\documentclass{ws-ijmpcs}

\usepackage{hyperref}
\usepackage[superscript]{cite}

\begin{document}

\newcommand{\peii}{$\pi_{e2}$}
\newcommand{\peiig}{$\pi_{e2\gamma}$}
\newcommand{\peiii}{$\pi_{e3}$}
\newcommand{\VmA}{$V$$-$$A$}

\markboth{D. Po\v{c}ani\'c \& al.}{Rare muon and pion decays}

%
\catchline{}{}{}{}{}
%
 
\title{NEW RESULTS IN RARE ALLOWED MUON AND PION
  DECAYS
      } 

\newcommand{\uva}{Inst.\ of Nuclear and Particle Physics, University
             of Virginia,  Charlottesville, VA 22904, USA} 
\newcommand{\dubna}{Joint Institute for Nuclear Research, RU-141980 Dubna,
                 Russia} 
\newcommand{\PSI}{Paul Scherrer Institut, Villigen PSI, CH-5232, Switzerland}
\newcommand{\IRB}{Institut Rudjer Bo\v{s}kovi\'c, HR-10000 Zagreb,
                 Croatia} 
\newcommand{\swierk}{NCBJ National Centre for Nuclear Research,
                  Otwock, Poland} 
\newcommand{\tbilisi}{Institute for High Energy Physics, Tbilisi State
                 University, GUS-380086 Tbilisi, Georgia} 
\newcommand{\unizh}{Physik-Institut, Universit\"at Z\"urich, CH-8057
                 Z\"urich, Switzerland}

\author{D.~PO\v{C}ANI\'C$^1$\footnote{Corresponding author:
          pocanic@virginia.edu}, 
        E.~MUNYANGABE$^1$\footnote{Present address:  Ngali Institute Ltd., 
          Kigali, Rwanda },    
        M.~BYCHKOV$^1$,      
        L.P.~ALONZI$^1$\footnote{Present address: CENPA, University of
          Washington, Seattle, WA 98195, USA},
        V.A.~BARANOV$^2$,   
        W.~BERTL$^3$,        
        Yu.M.~BYSTRITSKY$^2$, 
        E.~FRLE\v{Z}$^1$,    
        V.A.~KALINNIKOV$^2$, 
        N.V.~KHOMUTOV$^2$,   
        A.S.~KORENCHENKO$^2$, 
        S.M.~KORENCHENKO$^2$,  
        M.~KOROLIJA$^4$,      
        T.~KOZLOWSKI$^5$,     
        N.P.~KRAVCHUK$^2$,    
        N.A.~KUCHINSKY$^2$,   
        M.C.~LEHMAN$^1$,      
        D.~MEKTEROVI\'C$^4$,   
        D.~MZHAVIA$^{2,6}$\footnote{Deceased},    
        A.~PALLADINO$^{1,3}$\footnote{Present address: INFN Frascati, Italy},
        P.~ROBMANN$^7$,       
        A.M.~ROZHDESTVENSKY$^2$, 
        I.~SUPEK$^4$,          
        P.~TRU\"OL$^7$,        
        Z.~TSAMALAIDZE$^6$,    
        A.~VAN~DER~SCHAAF$^7$, 
        B.~ VANDEVENDER$^1$\footnote{Present address: Pacific
          Northwest National Laboratory, Richland, WA 99354, USA},\quad
        E.P.~VELICHEVA$^2$,    
        V.P.~VOLNYKH$^2$       
           }


\address{$^1$\uva \\
         $^2$\dubna \\
         $^3$\PSI \\
         $^4$\IRB  \\
         $^5$\swierk \\
         $^6$\tbilisi \\
         $^7$\unizh}

\maketitle

\begin{history}
\received{30 December 2013}
\revised{30 December 2013}
\end{history}

\begin{abstract}
Simple dynamics, few available decay channels, and highly controlled
radiative and loop corrections, make pion and muon decays a sensitive
means of exploring details of the underlying symmetries.  We review the
current status of the rare decays:
  $\pi^+ \to e^+\nu$ (\peii),
  $\pi^+ \to e^+\nu\gamma$ (\peiig),
  $\pi^+\to \pi^0 e^+ \nu$ (\peiii), and
  $\mu^+ \to e^+\nu\bar{\nu}\gamma$.  
For the latter we report new preliminary values for the branching ratio
  $B(E_{\gamma}>10\,\text{MeV},\theta_{e\gamma}>30^{\circ}) = 
   4.365\,(9)_{\text{stat}}\,(42)_{\text{syst}}  \times 10^{-3}$, 
and the decay parameter 
   $\bar{\eta} = 0.006\,(17)_{\text{stat}}\,(18)_{\text{syst}}$, both in
excellent agreement with standard model predictions.
We review recent measurements, particularly by the PIBETA and PEN
experiments, and near-term prospects for improvement.  These and other
similar precise low energy studies complement modern collider results
materially.

\end{abstract}

\ccode{PACS numbers: 13.20.Cz, 13.35.Bv 14.40.Be}

\section{The physics of rare muon and pion decays}

Muon decay, a pure leptonic electroweak process, serves a special role
in the standard model (SM) because it calibrates the strength of the
weak coupling.  Its precise theoretical description, via the so-called
Michel parameters\cite{Mic50}, positions it uniquely to provide
constraints on possible contributions outside the \VmA\ standard
electroweak model.  Below we discuss new results on the radiative muon
decay $\mu^+ \to e^+ \nu_{e} \bar{\nu}_\mu\gamma$, or RMD, the only
process that gives access to the decay parameter $\bar{\eta}$.

Pion decays constrain the SM in different ways, ever since providing
early evidence for the \VmA\ nature of the weak interaction through the
$\sim$$10^5$ helicity suppression of direct $\pi \to e$ decay
compared to the $\pi\to\mu\to e$ sequence.  In recent decades the
theoretical treatment of pion decays has reached unprecedented levels of
precision.  Of particular interest are the \peii, \peiig, and
\peiii\ decay channels. \\[3pt]
\noindent 
  \textbf{\boldmath(a) $\pi\to e\nu$ decay (\peii)}

A series of calculations have refined the SM description to a precision
better than a part in $10^4$: \vspace*{-6pt}
  \begin{equation}
     \left(R_{e/\mu}^{\pi}\right)^{\text{SM}} = 
       \frac{\Gamma(\pi \to e\bar{\nu}(\gamma))}
          {\Gamma(\pi \to  \mu\bar{\nu}(\gamma))}\bigg|_{\text{calc}} =
 \begin{cases}
    1.2352(5) \times 10^{-4} & \text{Ref.~\refcite{Mar93},} \\
    1.2354(2) \times 10^{-4} & \text{Ref.~\refcite{Fin96},} \\
    1.2352(1) \times 10^{-4} & \text{Ref.~\refcite{Cir07},} \\[-6pt]
   \end{cases}
\end{equation}
where $(\gamma)$ indicates that radiative decays are included.
Meanwhile, experimental precision lags behind the theoretical
one by almost two orders of magnitude:
$   \left(R_{e/\mu}^{\pi}\right)^{\text{exp}} = 
       1.230(4) \times 10^{-4}\,.
$\cite{PDG12}
Because of the large helicity suppression of the \peii\ decay, its
branching ratio is highly susceptible to small non-\VmA\  contributions
from new physics, making this decay a particularly suitable subject of
study, as discussed in, e.g.,
Refs.~\citen{Sch81,Sha82,Loi04,Ram07,Cam05,Cam08}.  This prospect
provides the primary motivation for the ongoing PEN\cite{PENweb} and
PiENu\cite{Agu09} experiments.  Of the possible ``new physics''
contributions in the Lagrangian, \peii\ is directly sensitive to the
pseudoscalar one.  At the precision of $10^{-3}$, $R_{e/\mu}^\pi$ probes
the pseudoscalar and axial mass scales up to 1,000\,TeV and 20\,TeV,
respectively\cite{Cam05,Cam08}.  For comparison, CKM unitarity and
precise measurements of several superallowed nuclear beta decays
constrain the non-SM vector contributions to 20\,TeV, and scalar to
10\,TeV \cite{PDG12}.  Although scalar interactions do not directly
contribute to $R_{e/\mu}^\pi$, they can do so through loop
effects, resulting in sensitivity to new scalar interactions up to
60\,TeV \cite{Cam05,Cam08}.  The subject was recently reviewed in
Ref.~\refcite{Bry11}.  Alternatively, $(R_{e/\mu}^{\pi})^{\text{exp}}$
provides limits on masses of certain SUSY partners\cite{Ram07}, and on
neutrino sector anomalies\cite{Loi04}. \\[3pt]
\noindent 
  \textbf{\boldmath(b) Radiative $\pi\to e\nu\gamma$ decay (\peiig)}

Thanks to the radiated photon, the \peiig\ decay provides information on
the structure of the pion.  Two sets of amplitudes contribute to RPD:
the inner-bremsstrahlung, IB, fully described by QED, and the
structure-dependent, SD.  Standard \VmA\ electroweak theory requires
only two pion form factors, $F_A$, axial vector, and $F_V$, vector, to
describe the SD amplitude; the CVC hypothesis fixes $F_V = 0.0259(9)$.
In a series of measurements at the Paul Scherrer Institute (PSI), the
PIBETA project, discussed below, has improved the experimental precision
of the \peiig\ differential branching ratio as well as $F_A$ and $F_V$
by over an order of magnitude, and set new limits on the putative tensor
term, $F_T$.\cite{Frl04a,Byc09}. \\[9pt]
\noindent 
  \textbf{\boldmath(c) Pion beta decay:
    $\pi^+\to\pi^0 e^+\nu$ (\peiii)} 

The extremely rare, ${\cal O}(10^{-8})$, pure vector $0^-\to 0^-$ pion
beta decay is the theoretically cleanest process to measure the
Cabibbo-Kobayashi-Maskawa (CKM) quark mixing matrix element $V_{ud}$
and test quark-lepton universality.  The PIBETA project has improved
the experimental precision of the branching ratio by an order of
magnitude\cite{Poc04}.  This result represents the most stringent test
of CVC and Cabibbo universality in a meson.  The urgency of a further
improvement was considerably reduced by the BNL E865
result\cite{She03} and the subsequent renormalization of $V_{us}$ that
removed a longstanding $2-3\sigma$ shortfall in CKM matrix unitarity.

\section{Experimental method: the PEN and PIBETA projects}

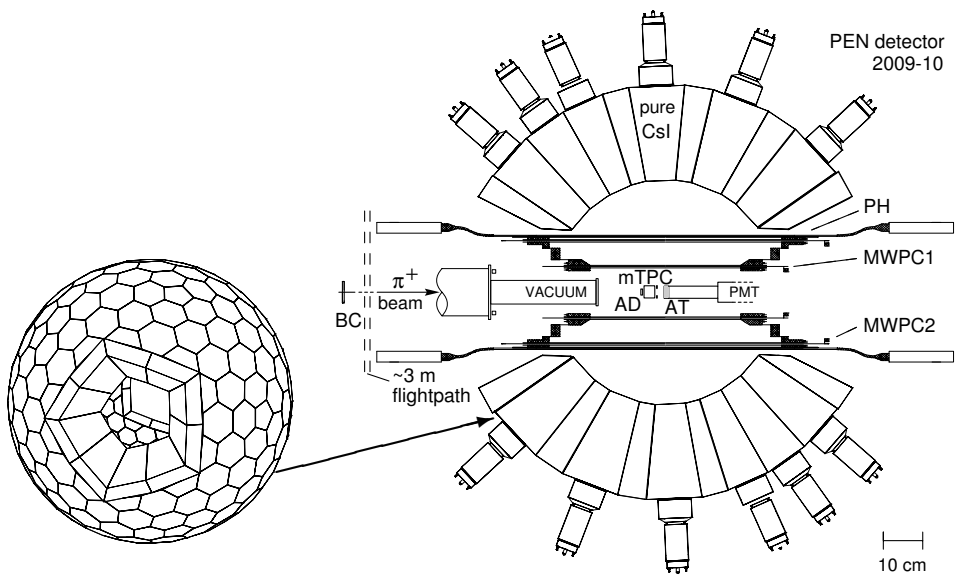
\begin{figure}[!b]
  
  \parbox[b]{0.3\linewidth}{
                \includegraphics[width=\linewidth]{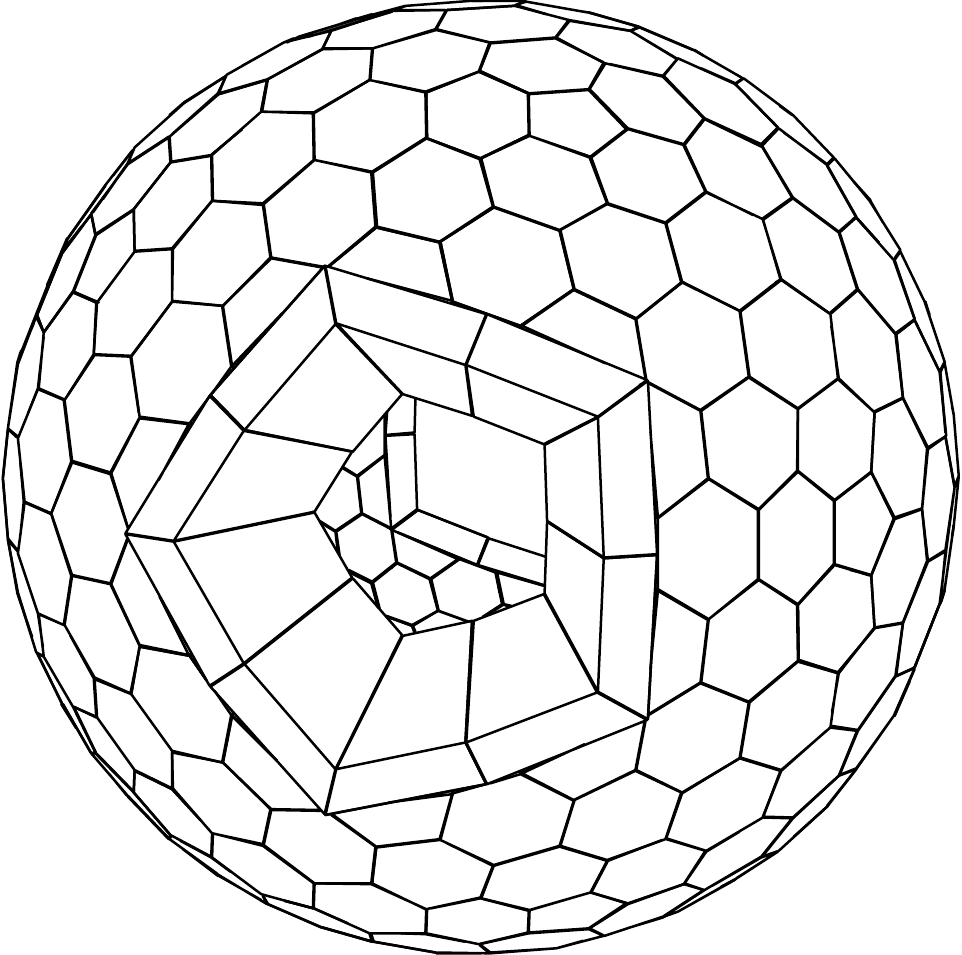}\\ 
  \begin{picture}(0,0)
    \thicklines
    \put(102,40){\vector(4,1){82}}
  \end{picture}
                           }
  \hspace*{\fill}
  \parbox[b]{0.66\linewidth}{
                \includegraphics[width=\linewidth]{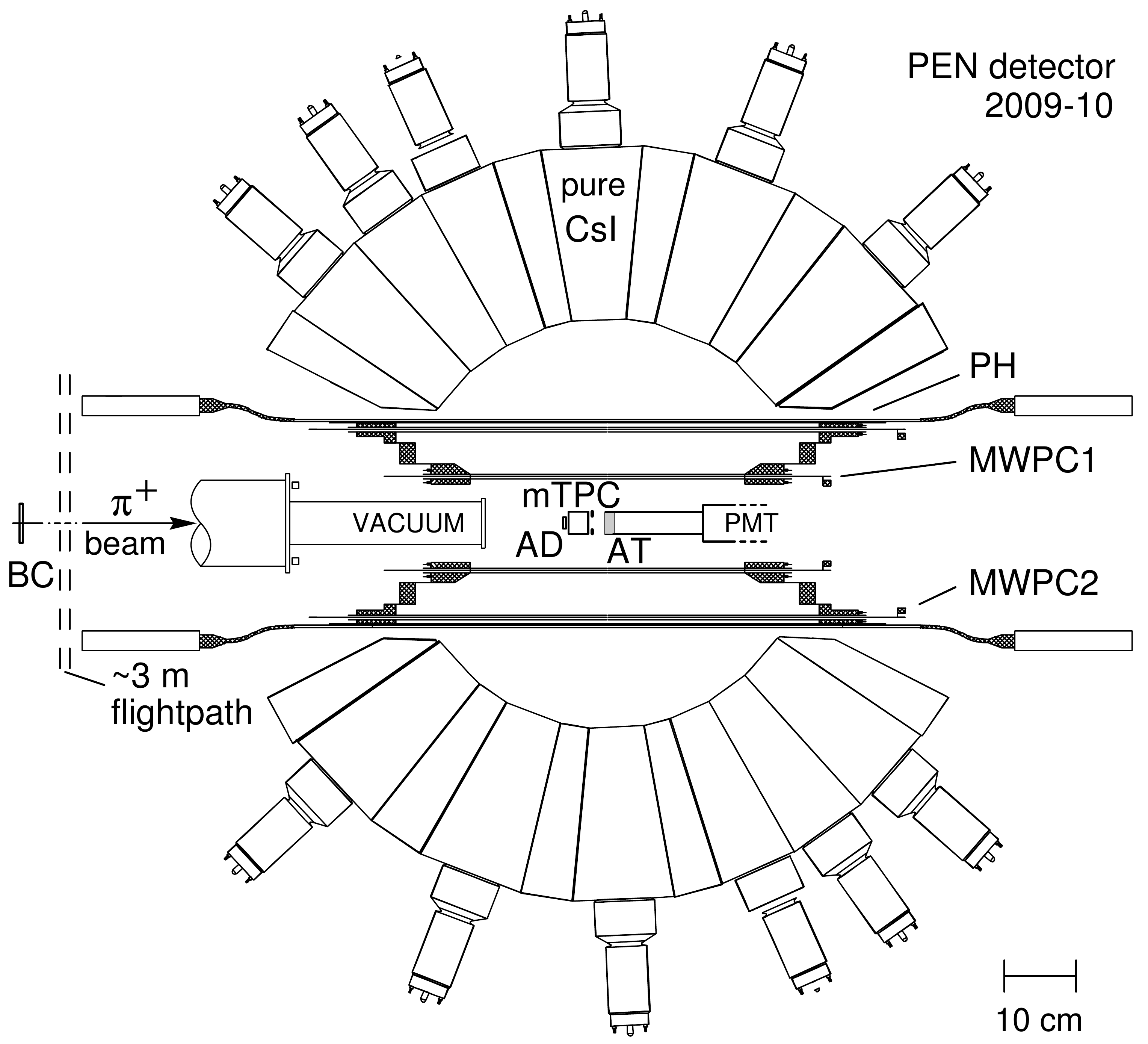}}
     \caption{Schematic cross section of the
      PIBETA/PEN apparatus, shown in the 2009 PEN 
      configuration, with its main components: beam entry with the
      upstream beam counter (BC), 5\,mm thick active degrader (AD),
      mini time projection chamber (mTPC) followed by a passive Al
      collimator, and active target (AT), cylindrical multiwire
      proportional chambers (MWPC's), plastic hodoscope (PH) detectors
      and photomultiplier tubes (PMT's), 240-element pure CsI
      electromagnetic shower calorimeter and its PMT's.  BC, AD, and
      PH detectors are made of plastic scintillator.}
      \label{fig:xsect} 
\end{figure}

The PEN\cite{PENweb} and its predecessor PIBETA\cite{PBweb}
experiments were designed to optimize the detection of pion and muon
decays at rest.  The apparatus, shown in Fig.~\ref{fig:xsect},
comprises an active target with beam tracking, two concentric
cylindrical multiwire proportional chambers, a thin 20-element
hodoscope, and a $\sim 3\pi$\,sr electromagnetic calorimeter made of
pure CsI.  The apparatus and its performance are described in detail in
Ref.~\refcite{Frl04b}.  PIBETA focused on the \peiii, \peiig\ and RMD
decays in four long runs between 1999 and 2004.  PEN, which started in
2007 and has conducted its main data runs in 2008--10, is focused on
\peii, but will produce important updates to the PIBETA \peiig\ and RMD
results.  Key to both experiments is the simultaneous precise treatment
of all of these decay processes, along with the dominant $\pi\to\mu\to
e$ chain, as each presents significant backgrounds to the others.

\section{\boldmath Radiative muon decay: $\mu^+\to e^+\nu\bar{\nu}\gamma$}

\begin{figure}[!htb]
  \parbox{0.49\linewidth}{
    \includegraphics[width=\linewidth]{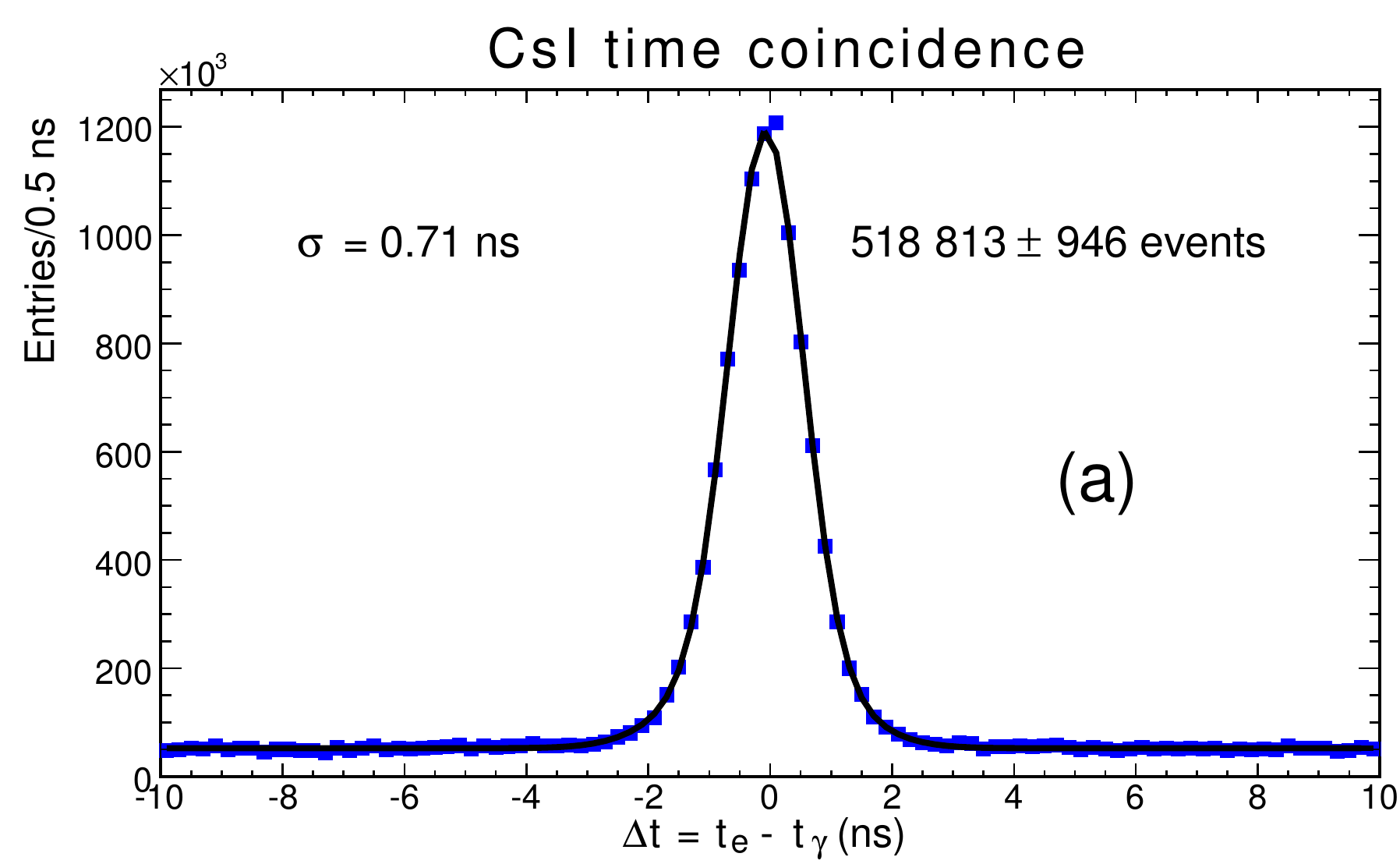}
                         }
  \parbox{0.49\linewidth}{
  \includegraphics[width=\linewidth]{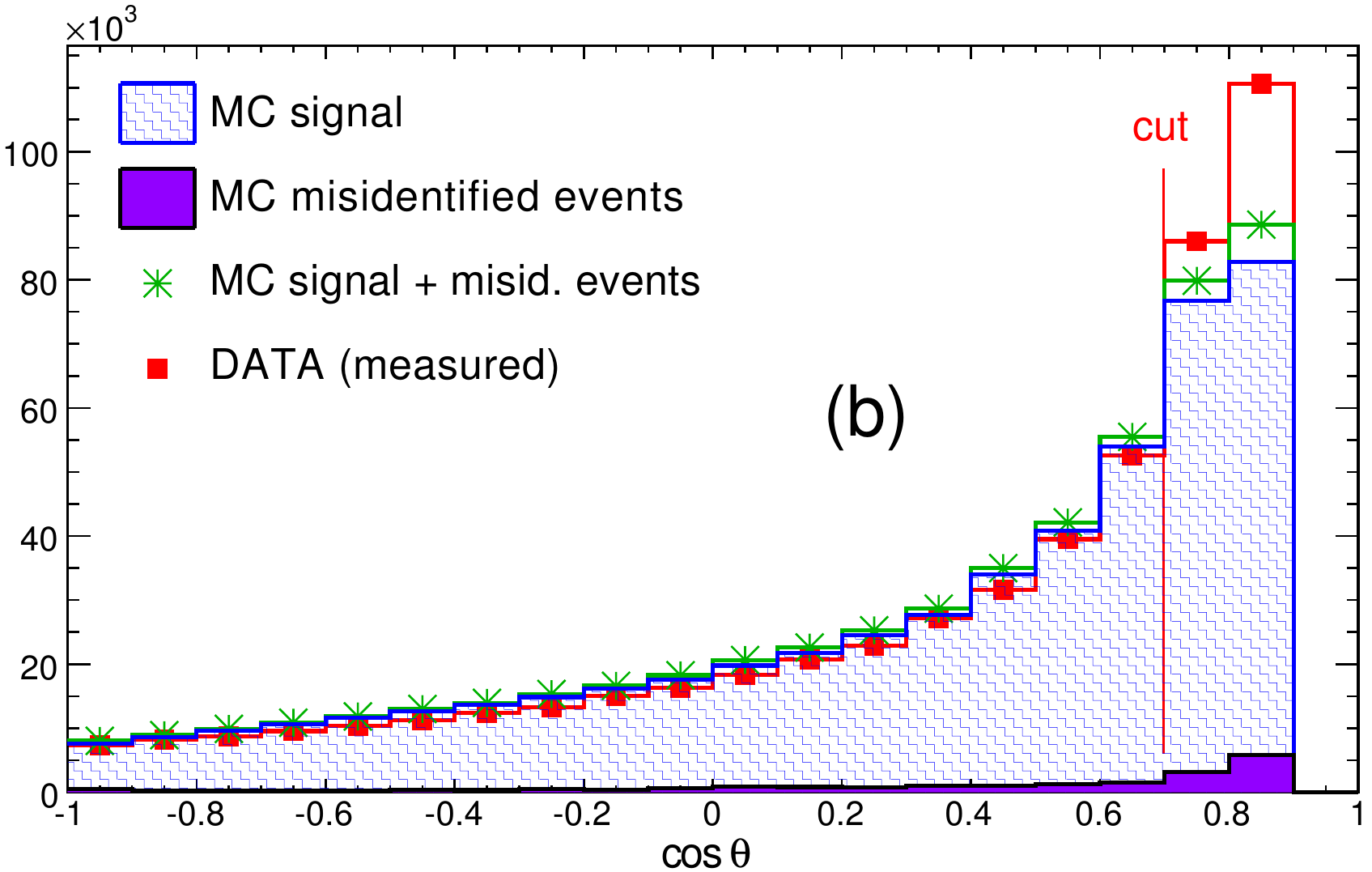} 
                         }
  \parbox{0.49\linewidth}{
    \includegraphics[width=\linewidth]{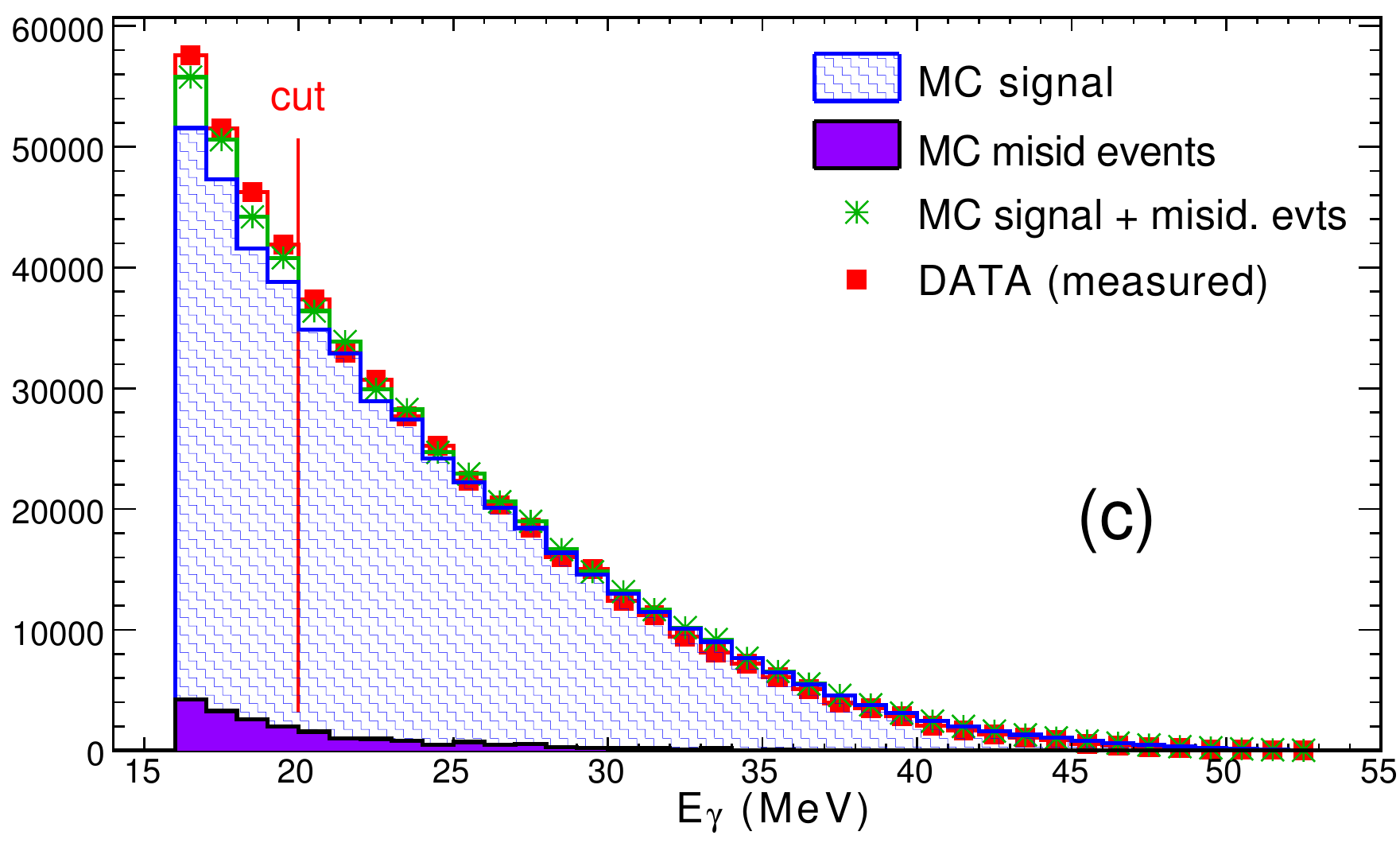}
                         }
  \parbox{0.49\linewidth}{
  \includegraphics[width=\linewidth]{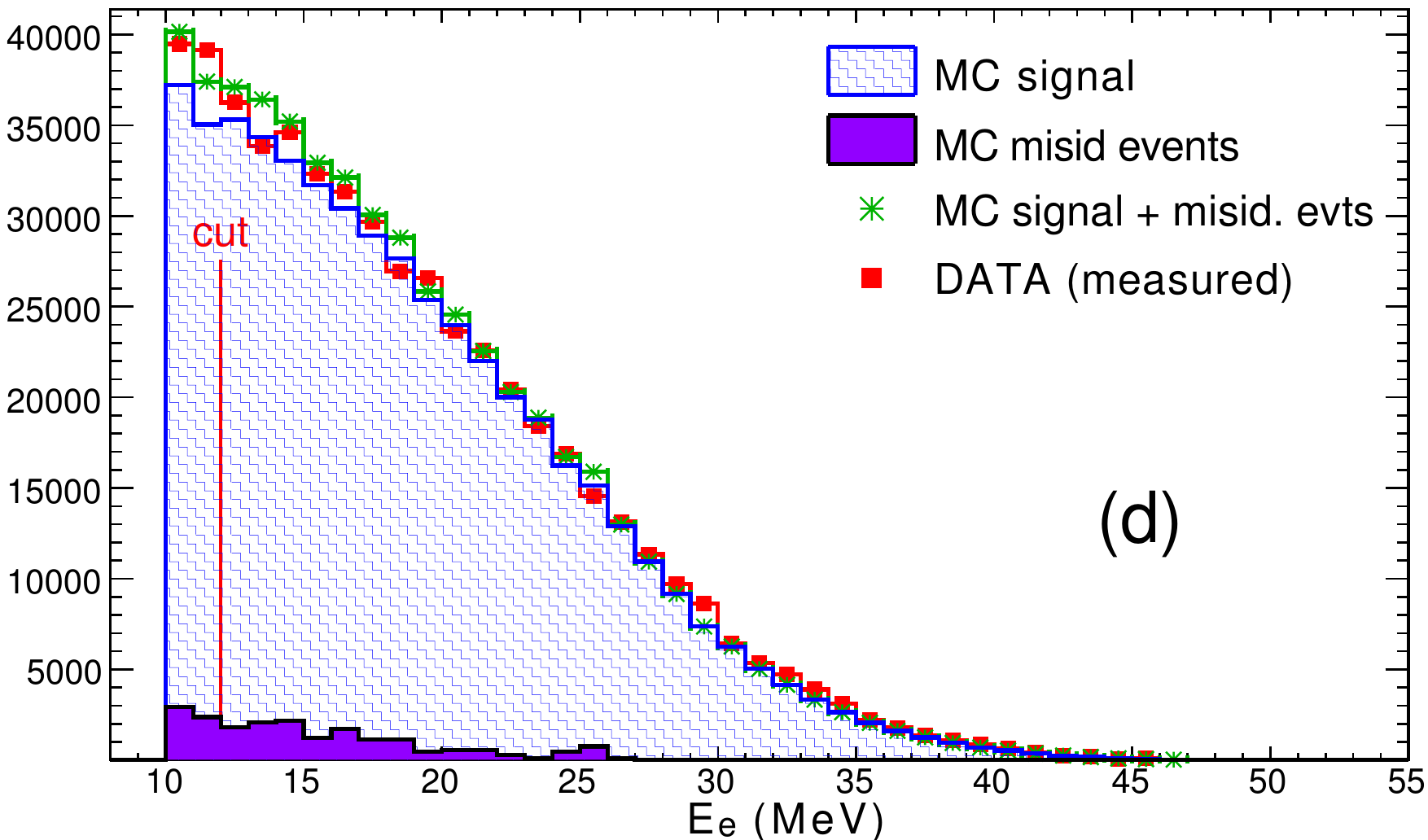}
                         }
  \caption{Measured and simulated RMD distributions of: (a) $\Delta
    t_{e\gamma}$, (b) $\cos\theta_{e\gamma}$, (c) $E_\gamma$, and (d)
    $E_{e^+}$.  Also shown are the misidentified Monte Carlo events
    (split-off secondary neutral showers), as well as bounds of cuts
    applied in the branching ratio analysis.
    \label{fig:rmd}}
\end{figure}

We have recently analyzed a set of $\sim$0.5\,M  RMD events; the
relevant measured and Monte Carlo simulated spectra, including
backgrounds, are shown in Fig.~\ref{fig:rmd}, showing excellent
agreement within the design acceptance of the spectrometer.  The
analysis yields a preliminary branching ratio for $E_\gamma >
10$\,MeV, and $\theta_{e\gamma} > 30^\circ$:
\begin{equation}
  B^{\text{exp}}  =
      4.365\,(9)_{\text{stat}}\,(42)_{\text{syst}}  \times 10^{-3}\,,
   \label{eq:B-RMD}
\end{equation}
which represents a 29-fold improvement in precision over the previous
result, and is in excellent agreement with the SM value: $B^{\text{SM}}
= 4.342\,(5)_{\text{stat-MC}} \times 10^{-3}$.  Minimum-$\chi^2$
analysis of the most sensitive data subset (with roughly balanced
systematic and statistical uncertainties) yields a preliminary value for
the $\bar{\eta}$ parameter ($\bar{\eta}^{\text{SM}}\equiv 0$):
\begin{equation}
  \bar{\eta} =  0.006\,(17)_{\text{stat}}\,(18)_{\text{syst}},\qquad
      \text{or} \qquad \bar{\eta}  < 0.028 \quad (68\% \text{CL})\,,
\end{equation}
a 4-fold improvement over previous limits.\cite{Eic84} Details of this
analysis, including a comprehensive discussion of the uncertainties, are
given in Refs.~\refcite{Mun12} and \refcite{Van05}.

\section{Prospects for the near future}

During the 2008-10 production runs the PEN experiment accumulated some
23\,M $\pi\to e\nu$, and $>150$\,M $\pi\to\mu\to e$ events, as well as
significant numbers of pion and muon radiative decays.  A
comprehensive blinded maximum likelihood analysis is under way to
extract a new experimental value of $R_{e/\mu}^{\pi}$.  The PEN goal
is $\Delta R/R \simeq 5\times 10^{-4}$.  The competing PiENu
experiment at TRIUMF has a similar precision goal.  The near to medium
future will thus bring about a substantial improvement in the limits
on $e$-$\mu$ lepton universality, and the attendant SM limits.

Once completed, analysis of the PEN \peiig\ data is expected to yield
improvements in the SD$^-$ structure dependent amplitude, which
constrains $F_V-F_A$, thus improving on the PIBETA result for
$F_V$\cite{Byc09}.  In a similar vein, analysis of the PEN $\mu^+\to
e^+\nu\bar{\nu}\gamma$ data is expected to improve the present value
of the Michel parameter $\bar{\eta}$, leading to a new global analysis
of the fundamental weak couplings.

\section*{Acknowledgments}

This work has been supported by grants from the US National Science
Foundation (most recently PHY-1307328), the Paul Scherrer Institute,
and the Russian Foundation for Basic Research (Grant 13-02-00745А).



\begin{thebibliography}{00}  

\bibitem{Mic50} L. Michel, \textit{Proc. Phys.\ Soc.\ London Sect.\ A}
  \textbf{63}, 514 (1950); C. Bouchiat and L. Michel,
  \textit{Phys. Rev.} \textbf{106}, 170 (1957); T. Kinoshita and
  A. Sirlin, \textit{ibid.} \textbf{108}, 844 (1957).

\bibitem{Mar93} W.J. Marciano and A. Sirlin,
    \textit{Phys.\ Rev.\ Lett.}\ {\bfseries 71}, 3629 (1993).

\bibitem{Fin96} M. Finkemeier, \textit{Phys.\ Lett.\ B} {\bfseries 387},
  391 (1996).

\bibitem{Cir07} V. Cirigliano and I. Rosell, \textit{Phys.\ Rev.\ Lett.}
  {\bfseries 99}, 231801 (2007).

\bibitem{PDG12} J. Beringer \textit{et al.} (Particle Data Group),
  \textit{Phys.\ Rev.\ D} \textbf{86}, 010001 (2012).

\bibitem{Sch81} R.E.~Schrock, \textit{Phys.\ Rev.\ D} \textbf{24}, 5
  (1981).

\bibitem{Sha82} O.U.~Shanker, \textit{Nucl.\ Phys.} \textbf{B204}, 375
  (1982). 

\bibitem{Loi04} W.~Loinaz,  N.~Okamura,  S.~Rayyan \textit{et al.},
  \textit{Phys.\ Rev.\ D} \textbf{70}, 113004 (2004).

\bibitem{Ram07}M.J.~Ramsey-Musolf, S.~Su, and S.~Tulin,
  \textit{Phys.\ Rev.\ D} \textbf{76}, 095017 (2007).


\bibitem{Cam05} B.A.~Campbell and D.W.~Maybury, \textit{Nucl.\ Phys.}
  \textbf{B709}, 419-439 (2005).

\bibitem{Cam08} B.A.~Campbell and Ahmed Ismail, arXiv 0810.4918, 2008.

\bibitem{Bry11} D.~Bryman, WJ.~Marciano \textit{et al.},
  \textit{Annu.\ Rev.\ Nucl.\ Part.\ Sci.} \textbf{61}, 331 (2011).

\bibitem{PENweb}  \url{http://pen.phys.virginia.edu/} and links therein.

\bibitem{Frl04a} E.~Frle\v{z}, D.~Po\v{c}ani\'c, V.A.~Baranov
  \textit{et al.}, \textit{Phys.\ Rev.\ Lett.} \textbf{93}, 181804
  (2004).

\bibitem{Byc09} M.~Bychkov, D.~Po\v{c}ani\'c  
  \textit{et al.}, \textit{Phys.\ Rev.\ Lett.} \textbf{103}, 181803
  (2009).

\bibitem{Poc04} D.~Po\v{c}ani\'c, E.~Frle\v{z}  
  \textit{et al.}, \textit{Phys.\ Rev.\ Lett.} \textbf{93}, 181803
  (2004). 

\bibitem{She03} A.~Sher, R.~Appel, G.S.~Atoyan
  \textit{et al.}, \textit{Phys.\ Rev.\ Lett.} \textbf{91}, 261802 (2003). 

\bibitem{PBweb} \url{http://pibeta.phys.virginia.edu/} and links
  therein. 

\bibitem{Agu09} A. Aguilar-Arevalo, M. Blecher \textit{et al.},
  \textit{Nucl.\ Instrum.\ Methods A} \textbf{609}, 102 (2009).

\bibitem{Frl04b} E.~Frle\v{z}, D.~Po\v{c}ani\'c 
  \textit{et al.}, \textit{Nucl.\ Instrum.\ Methods A} \textbf{526}, 300
  (2004).

\bibitem{Eic84} W.~Eichenberger, R.~Engfer, and A.~van~der~Schaaf,
  \textit{Nucl.\ Phys.} \textbf{A412}, 523 (1984).

\bibitem{Mun12}  E.~Munyangabe, Ph.D. thesis, University of Virginia,
  2012. 

\bibitem{Van05} B.A.~VanDevender, Ph.D. thesis, University of Virginia, 2005.



\end{thebibliography}
\end{document}